\documentclass[pdflatex,sn-mathphys-num]{sn-jnl}

\usepackage{graphicx}%
\usepackage{multirow}%
\usepackage{amsmath,amssymb,amsfonts}%
\usepackage{amsthm}%
\usepackage{mathrsfs}%
\usepackage[title]{appendix}%
\usepackage{xcolor}%
\usepackage{textcomp}%
\usepackage{manyfoot}%
\usepackage{booktabs}%
\usepackage{algorithm}%
\usepackage{algorithmicx}%
\usepackage{algpseudocode}%
\usepackage{listings}%
\usepackage{bm}%

\theoremstyle{thmstyleone}%

\theoremstyle{thmstyletwo}%

\theoremstyle{thmstylethree}%

\newcommand{\dd}{\mathrm{d}}
\newcommand{\ii}{\mathrm{i}}
\newcommand{\rstar}{r_{\ast}}
\newcommand{\Veff}{V_{\mathrm{eff}}}

\begin{document}
	
	\title[Spectral densification and macroscopic phase delay of gravitational echoes from exotic compact objects]{Spectral densification and macroscopic phase delay of gravitational echoes from exotic compact objects}
	
	\author*[1]{\fnm{Corentin} \sur{Guigot}}\email{cguigot@cesi.fr}
	
	\affil*[1]{\orgdiv{CESI LINEACT}, \orgname{Campus CESI}, \orgaddress{\street{24, Le Paquebot - CS 60133}, \city{Saint-Nazaire}, \postcode{44600}, \country{France}}}
	
	\abstract{Gravitational-wave echoes from Exotic Compact Objects (ECOs) provide an observable probe for horizon-scale physics. Standard phenomenological models for these signals typically assume a constant Free Spectral Range, relying on the geometric optics approximation. In this work, we demonstrate that wave dispersion at the photon sphere induces a systematic deviation from this assumption, manifesting instead as a hyperbolic spectral densification. By employing an analytical framework based on the Riccati equation and macroscopic impedance mapping, we extract the spectrum of these high-finesse resonances without semi-classical approximations. We characterize the structural transition from the eikonal geometric asymptote ($\ell \gg 1$) down to the wave-tunneling dominated quadrupolar mode ($\ell=2$). In this wave-dominated regime ($\ell \in [2, 10]$), the macroscopic deviation from the semi-classical limit is governed by a phenomenological $\mathcal{L}^{-3/2}$ inverse power law. Finally, we show that this macroscopic densification isolates the structural dispersion of the external spacetime, decoupled from the boundary microphysics, provided the membrane phase shift is frequency-independent.}
	
	
	\maketitle
	
	\section{Introduction}
	
	The first direct detection of gravitational waves \cite{Abbott2016} initiated the era of precision tests of General Relativity and black hole spectroscopy \cite{Baibhav2019}. While the ringdown phase of GW150914 appears consistent with the classical Kerr hypothesis \cite{Isi2019, LIGOScientific2021}, the resolving power of current interferometers does not unequivocally rule out compact remnants that deviate from the black hole paradigm \cite{Cardoso2019}. Such Exotic Compact Objects (ECOs) (including gravastars \cite{Mazur2004}, fuzzballs \cite{Mathur2005}, wormholes \cite{Damour2007}, or boson stars \cite{Liebling2012}) aim to resolve the information paradox by replacing the event horizon with a physical structure at or near the Planck scale.
	
	Dynamically, this modification transforms the near-horizon geometry into a macroscopic resonant cavity. Radiation confined between the inner reflective surface and the outer centrifugal barrier effectively circles the compact object, forming high-finesse quasi-bound states. Mathematically, the propagation of these gravitational perturbations through the Regge-Wheeler potential is isomorphic to the trapping of electromagnetic waves in dielectric micro-resonators. These states therefore behave as macroscopic Whispering Gallery Modes (WGMs) \cite{Kippenberg2011, Gorodetsky1996, Righini2019, Guigot2021, Guigot2024}. Just as optical WGMs exhibit phase shifts and structural dispersion  dictated by the boundaries of the confining medium, the frequency comb of these gravitational WGMs encodes the exact macroscopic phase delays induced by the local spacetime curvature. These localized resonances leak out periodically, producing post-merger gravitational-wave echoes, a phenomenological signature first modeled for ultracompact constant-density stars \cite{Kokkotas1997, Cardoso2016, Cardoso2017}. Although tentative claims of such signals exist \cite{Abedi2017}, their statistical significance remains contested \cite{Westerweck2018, LIGOScientific2021_TGR}. Robust extraction via matched filtering requires extreme phase coherence, demanding highly accurate theoretical templates \cite{Berti2009, Konoplya2011}. Quantifying the underlying phase dispersion is thus a prerequisite for the next generation of space-based and terrestrial observatories, such as LISA \cite{AmaroSeoane2017} and the Einstein Telescope \cite{Maggiore2020}.
	
	Computing these signatures poses a specific phenomenological challenge. Standard models often rely on the geometric optics approximation \cite{Mark2017}, which assumes the Free Spectral Range (FSR) of the cavity is dictated purely by the light-crossing time ($\Delta\omega \approx \pi/L$). However, ECO potentials feature extended potential barriers that induce non-trivial wave dispersion at the photon sphere, affecting the accuracy of such approximations \cite{Schutz1985, Bueno2018}. While the semi-classical eikonal limit ($l \gg 1$) is structurally well understood, the transition down to the wave-tunneling dominated regime ($l=2, 3$), which constitutes the primary emission channel for binary coalescences, exhibits macroscopic phase delays that first-order analytical approximations do not fully capture.
	
	To quantify these dispersive effects across all angular momentum regimes, we exploit an analytical formalism establishing a direct link with stratified photonics \cite{Yeh2005,Vahala2003}. The application of the Riccati equation to Schwarzschild perturbations was pioneered by Chandrasekhar and Detweiler \cite{Chandrasekhar1975}, while the concept of discretizing the effective potential to study spectral properties was introduced by Nollert \cite{Nollert1996}. Building upon these foundational works, we implement a macroscopic impedance mapping. By integrating the gravitational impedance rather than the highly oscillatory wavefunction, this framework naturally isolates the dispersive phase delays without semi-classical approximations. This approach characterizes the breakdown of the geometric box assumption, identifying a hyperbolic scaling law for the spectral densification and isolating the $\mathcal{L}^{-3/2}$ phenomenological inverse power law that governs the macroscopic wave-tunneling deviation.
	
	The rest of this paper is organized as follows. In Sec.~II, we introduce the impedance mapping derived from the Regge-Wheeler equation and establish the phenomenological framework for spectral extraction. In Sec.~III, we evaluate the deviation from the geometric FSR, establish the scaling laws governing the spectral densification from the eikonal to the wave regime, and validate the structural invariance of this dispersion. Finally, we discuss the physical domain of validity of the static model and summarize our conclusions in Sec.~IV.

	\section{Formalism: The $N$-Layer Resonator Analogy and Bilateral Matching}
	
	The global morphology of gravitational echoes and the associated resonance spectrum have been extensively characterized using S-matrix scattering amplitudes and transfer-function formalisms \cite{Mark2017, Conklin2019}. In these frameworks, the resonance condition naturally emerges from the poles of the transfer function. However, isolating the precise frequency-dependent phase delay induced solely by the centrifugal barrier remains analytically challenging at low multipoles ($\ell=2,3$). Our Riccati impedance mapping is specifically designed to complement these transfer-function methods by providing a high-resolution numerical extraction of this dispersive delay.
	
\subsection{From Wavefunction to Impedance Mapping}

The radial dynamics of gravitational (spin-$2$) perturbations $\Psi(\rstar)$ in a static, spherically symmetric Schwarzschild background of mass $M$ is governed by the Regge-Wheeler master equation:
\begin{equation}
	\frac{\dd^2\Psi}{\dd\rstar^2} + \left[\omega^2 - \Veff(\rstar)\right]\Psi = 0.
	\label{eq:RW_formalism}
\end{equation}
Here, $\rstar$ is the standard tortoise coordinate defined by $\dd\rstar/\dd r = (1 - 2M/r)^{-1}$, and $\Veff$ is the effective Regge-Wheeler potential for axial perturbations with angular momentum $l$:
\begin{equation}
	\Veff(r) = \left(1 - \frac{2M}{r}\right) \left[ \frac{l(l+1)}{r^2} - \frac{6M}{r^3} \right].
	\label{eq:RW_potential}
\end{equation}
To formally isolate the dispersive phase dynamics, we introduce the gravitational impedance $Z(\rstar) = -\ii \Psi^{-1} \dd\Psi/\dd\rstar$. Substituting this ansatz into Eq.~(\ref{eq:RW_formalism}) transforms the linear second-order differential equation into a non-linear first-order Riccati equation, an approach first introduced for Schwarzschild perturbations by Chandrasekhar and Detweiler \cite{Chandrasekhar1975}:
\begin{equation}
	\ii \frac{\dd Z}{\dd\rstar} - Z^2 + \omega^2 - \Veff(\rstar) = 0.
	\label{eq:riccati_formalism}
\end{equation}
The analytical advantage of this mapping is fundamental: $Z(\rstar)$ translates the highly oscillatory behavior of $\Psi$ into a smoothly varying impedance profile. This framework explicitly isolates the macroscopic phase delays, bypassing the wave-separation ambiguities inherent to boundary-value problems in the deep tunneling regime \cite{Yeh2005}.

\begin{figure}[htbp]
		\centering
		\includegraphics{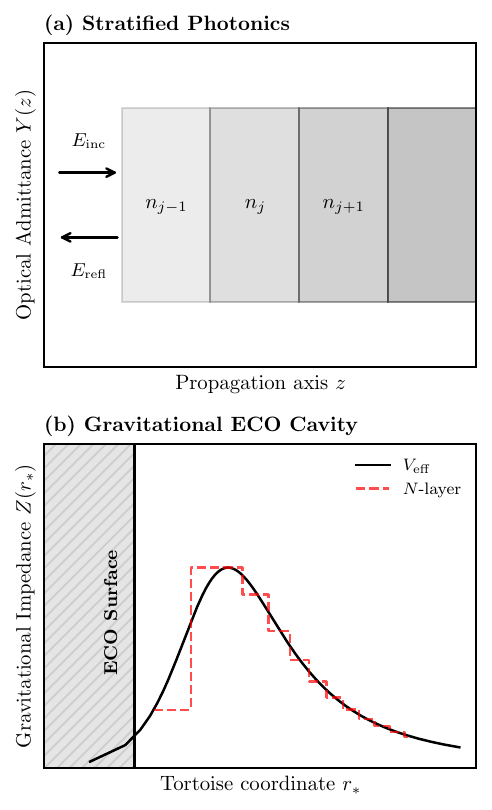}
		\caption{Schematic representation of the isomorphism between stratified photonics and the gravitational effective potential. (a) An electromagnetic wave propagating through a multilayer dielectric stack, parameterized by the optical admittance $Y(z)$. (b) Conceptual discretization of the Regge-Wheeler potential into an equivalent $N$-layer resonant cavity bounded by the perfectly reflecting ECO surface. The continuous spacetime dynamics are rigorously captured by mapping the gravitational impedance $Z(\rstar)$ across these homogeneous slabs.}
		\label{fig:analogy}
	\end{figure}

To physically ground this mathematical transformation, we establish a structural isomorphism between the exterior Schwarzschild geometry and a stratified dielectric medium, as illustrated in Figure~\ref{fig:analogy}. The continuous gravitational barrier $\Veff(\rstar)$ is remodeled as an $N$-layer optical resonator. In this paradigm, the highly oscillatory wave propagation through the curved geometry is strictly equivalent to the transmission of an electromagnetic wave through a multilayer stack, where the macroscopic gravitational impedance $Z(\rstar)$ serves as the exact dynamical analogue of the local optical impedance. This discrete impedance mapping is a standard, robust formalism for evaluating resonance conditions and spectral hybridization in optical micro-resonators \cite{Guigot2026}.
	
	\subsection{The Recursive $N$-Layer Propagator}
	
	Following the approach pioneered by Nollert to study spectral instabilities \cite{Nollert1996}, we discretize the continuous effective potential $\Veff(\rstar)$ into a sequence of $N$ homogeneous slabs of constant potential $V_j$ and width $\Delta \rstar$. The local wave vector within the $j$-th layer is $k_j = \sqrt{\omega^2 - V_j}$.
	
	Enforcing the continuity of the wavefunction and its derivative at each interface yields a recursive Möbius transformation. For a \textit{forward} propagation, the output impedance $Z_j$ is computed via:
	\begin{equation}
		Z_j = k_j \frac{Z_{j-1} + \ii k_j \tan(k_j \Delta \rstar)}{k_j + \ii Z_{j-1} \tan(k_j \Delta \rstar)}.
		\label{eq:propagator_forward}
	\end{equation}
	For a \textit{backward} propagation, the mapping requires a sign inversion due to the odd parity of the tangent function over $-\Delta \rstar$:
	\begin{equation}
		Z_{j-1} = k_j \frac{Z_j - \ii k_j \tan(k_j \Delta \rstar)}{k_j - \ii Z_j \tan(k_j \Delta \rstar)}.
		\label{eq:propagator_backward}
	\end{equation}
	This discrete mapping preserves the analytic structure of the wave propagation across all scattering regimes. In the tunneling regime ($V_j > \omega^2$), the wave vector becomes purely imaginary, and the trigonometric functions analytically transition to their hyperbolic counterparts, formally capturing the evanescent phase accumulation without introducing semi-classical matching artifacts.

\subsection{Boundary Conditions and Bilateral Phase Matching}

We model the ECO surface at $\rstar = R_{\text{wall}}$ as a perfectly reflecting Dirichlet boundary ($\Psi=0$), corresponding to $Z_{\text{wall}} \to \infty$. At spatial infinity, the outgoing Sommerfeld radiation condition imposes $Z_{\infty} = \omega$.

To preserve the integrity of the phase extraction and prevent the subdominant physical solution from being analytically overshadowed by the exponentially growing unphysical branch, we implement a bilateral matching technique at an intermediate boundary ($\rstar = 0$, corresponding to $r \approx 2.55M$, safely within the tunneling barrier). Specifically, the left domain is integrated \textit{forward} from the ECO membrane using Eq.~(\ref{eq:propagator_forward}), yielding $Z_{L}(\omega)$ at $\rstar=0$. Simultaneously, the right domain is integrated \textit{backward} from spatial infinity using Eq.~(\ref{eq:propagator_backward}), yielding $Z_{R}(\omega)$ at the same interface.

The discrete spectrum is isolated by finding the roots of the matching residual in the complex plane:
\begin{equation}
	Z_L(\omega) - Z_R(\omega) = 0.
	\label{eq:matching_condition}
\end{equation}

\textit{Asymptotic truncation and differential error cancellation.} The outgoing Sommerfeld radiation condition formally requires $Z_{\infty} = \omega$ as $\rstar \to +\infty$. However, the Schwarzschild effective potential decays slowly as a centrifugal tail $\sim 1/\rstar^2$. Enforcing a flat-space boundary condition at a finite spatial truncation (e.g., $\rstar = 100M$) introduces a systematic asymptotic phase error. Extending the spatial domain to arbitrary distances is analytically obstructed by the exponential divergence of the Lambert-W coordinate inversion ($\exp(\rstar/2M) \to \infty$). 

To mitigate this, we impose a first-order local wavevector matching: $Z_{\infty}(\omega) = \sqrt{\omega^2 - \Veff(r_{\infty})}$, which effectively absorbs the leading geometric tail at the analytical boundary $r_{\infty}$. While this does not substitute for hyperboloidal slicing, its residual impact on our results is mathematically neutralized by the differential nature of the observable. The spectral densification factor $K$ is derived from the Free Spectral Range ($\Delta\omega = \omega_{n+1} - \omega_n$). The residual asymptotic phase shift induces a nearly identical global drift on adjacent modes; therefore, the truncation error cancels out to the leading order during the subtraction. The macroscopic scaling laws extracted for the photon sphere dispersion are therefore robust against this finite-boundary approximation.

\textit{Numerical implementation and convergence stability.} To ensure the robustness of the root-finding procedure, the integration domain is uniformly partitioned into $N = 25,000$ constant-potential layers. The outer boundary condition is evaluated at $r_{*,\infty} = 200M$, and the bilateral matching interface is anchored at $r_{*} = 0$. The resonance extraction utilizes a two-step algorithm: a dense initial scan isolates the local minima of the matching residual, followed by a high-resolution refinement using Brent's method with an absolute tolerance of $10^{-12}$. Routine convergence tests establish that doubling the layer density or extending the outer truncation boundary to $r_{*,\infty} = 100M$ modifies the extracted real frequencies by less than $\mathcal{O}(10^{-6} M^{-1})$. This numerical variance is negligible compared to the macroscopic phase dispersion driving the reported densification scaling laws.

	\section{Results: Spectroscopy and Dispersive Densification}
	
	We apply the bilateral impedance framework to resolve the high-resolution resonance spectrum of a Schwarzschild-like ECO, utilizing the Lambert-W analytic inversion for the spatial coordinates to preserve the strictly conservative nature of the phase mapping.
	
	\subsection{Threshold Behavior and the Breakdown of the Geometric FSR}
	
	We characterize the resonant cavity by extracting the Free Spectral Range (FSR) between consecutive resonance modes. These trapped modes, supported by the external Schwarzschild potential and the internal ECO boundary, exhibit a morphology structurally analogous to whispering gallery phenomena, where waves are confined by the curvature of the effective potential rather than a physical dielectric boundary. For a standard flat-spacetime cavity of size $L$, the FSR is dictated purely by the geometric light-crossing time, $\Delta \omega_{\text{geom}} = \pi / L$. Our spectral analysis demonstrates a systematic macroscopic deviation from this geometric limit. 
	
\begin{figure}[htbp]
		\centering
		\includegraphics[width=\linewidth]{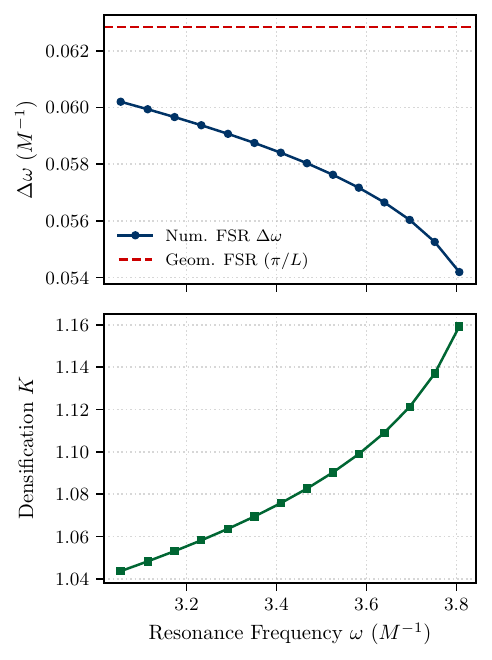}
		\caption{Spectral dispersion for an ECO membrane at $L=50M$ ($l=20$). The physical frequencies are correctly normalized to the standard $M=1$ convention ($\omega < \omega_{\max} \approx 3.93 M^{-1}$). The top panel isolates the extracted FSR dropping below the geometric asymptote ($\pi/L$), while the main panel illustrates the corresponding frequency-dependent densification factor $K(\omega)$.}
		\label{fig:spectrum}
	\end{figure}

	As depicted in Figure~\ref{fig:spectrum} for a reference cavity ($L=50M$), the dispersive FSR is not constant; it drops systematically below the geometric prediction. This structural deviation is quantified by defining the frequency-dependent spectral densification factor $K(\omega) = \Delta \omega_{\text{geom}} / \Delta \omega(\omega)$. The extraction reveals that $K$ departs from unity as the mode frequency approaches the classical trapping limit, directly demonstrating the limitations of simplified rigid-box models.
	
	Because this phase dispersion is deeply frequency-dependent near the barrier, extracting a representative scalar value requires a rigorous definition. We define the macroscopic densification factor $\langle K \rangle = \Delta \omega_{\text{geom}} / \langle \Delta \omega \rangle$, where $\langle \Delta \omega \rangle$ is the average modal spacing across the high-transmission leaky spectral window (the uppermost modes responsible for the early echoes).
	
\begin{figure}[htbp]
		\centering
		\includegraphics[width=0.8\linewidth]{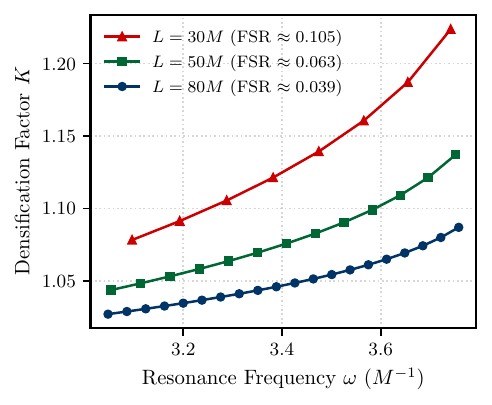}
		\caption{Parametric sweep of the spectral densification factor $K(\omega)$ for varying cavity sizes $L$. The macroscopic phase dispersion structurally intensifies for more compact ECOs.}
		\label{fig:fsr_analysis}
	\end{figure}

	Furthermore, a systematic evaluation across different cavity sizes ($L=30M, 50M, 80M$), illustrated in Figure~\ref{fig:fsr_analysis}, proves that this dispersive densification effect is structurally dependent on the membrane's position.
	
\subsection{Macroscopic Phase Delay and Regime Transition}

\begin{figure}[htbp]
	\centering
	\includegraphics[width=0.8\linewidth]{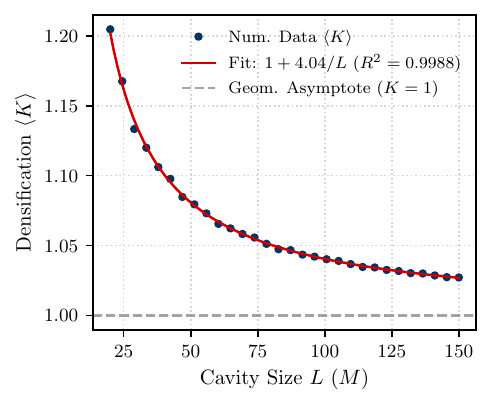}
	\caption{Mean densification factor $\langle K \rangle$ as a function of cavity size $L$. The extracted spectrum follows the macroscopic scaling dictated by the Wigner delay.}
	\label{fig:scaling_law}
\end{figure}

Analytically, the FSR is inversely proportional to the total round-trip time, $\tau_{\text{tot}} = 2L + \tau_{\text{delay}}$, where $\tau_{\text{delay}}$ is the Wigner phase delay \cite{Wigner1955} induced by the reflection at the Regge-Wheeler barrier. This yields a macroscopic phenomenological relation:
\begin{equation}
	\langle K \rangle(L) = 1 + \frac{C_l}{L},
	\label{eq:scaling_law}
\end{equation}
where $C_l = \langle \tau_{\text{delay}} \rangle / 2$ represents the frequency-averaged macroscopic Wigner delay over the leaky window. As shown in Figure~\ref{fig:scaling_law}, our numerical spectral data ($L \in [20M, 150M]$) is in excellent agreement with this $1/L$ macroscopic scaling ($R^2 > 0.997$).       

The physical origin of the densification constant $C_l$ is validated by comparing the extracted delay with analytical predictions derived from the Wentzel-Kramers-Brillouin (WKB) approximation and semi-classical scattering theory \cite{FordWheeler1959}. The semi-classical Wigner phase delay is given by the spatial integral between the ECO surface and the classical turning point $r_{\text{tp}}(\omega)$:
\begin{equation}
	\tau_{\text{delay}}(\omega) = 2 \int_{-L}^{r_{\text{tp}}(\omega)} \left[ 1 - \frac{\Veff(\rstar)}{\omega^2} \right]^{-1/2} \dd\rstar - 2L.
\end{equation}

Extracting a macroscopic, cavity-independent delay $\langle \tau_{\text{delay}} \rangle$ from this formalism presents a methodological challenge. As the mode frequency approaches the peak of the Regge-Wheeler barrier ($\omega \to \omega_{\text{crit}}$), the WKB integral develops a logarithmic singularity, predicting an infinite delay. However, modes deeply trapped in the potential well ($\omega \ll \omega_{\text{crit}}$) behave as standing waves in a flat box, diluting the dispersive signature of the photon sphere. Relying on a fixed number of modes near the barrier peak introduces a cavity-size dependence: for larger cavities (larger $L$), the spectral density increases, pushing the highest modes exponentially closer to the singular crest and artificially inflating the extracted constant $C_l$.

We neutralize this sampling artifact and ensure a scale-invariant comparison between the impedance mapping and the analytical WKB prediction by enforcing a macroscopic spectral averaging. We define a leaky spectral window between $85\%$ and $98\%$ of the barrier height ($\omega_{\text{crit}}$). This window avoids the breakdown of the first-order WKB approximation at the turning point while isolating the high-transmission modes responsible for early gravitational-wave echoes. The spectral densification constant is extracted by averaging the Free Spectral Range over the resonant comb enclosed within this fixed relative depth. Sensitivity tests establish that the extracted macroscopic delay is statistically robust against boundary variations of this domain; shifting the integration bounds down to $80\%$--$95\%$ preserves the mean densification factor up to $\mathcal{O}(10^{-5})$ precision, confirming that the extraction operates securely within a stable averaging plateau.

\begin{figure*}[t]
	\centering
	\includegraphics[width=0.85\textwidth]{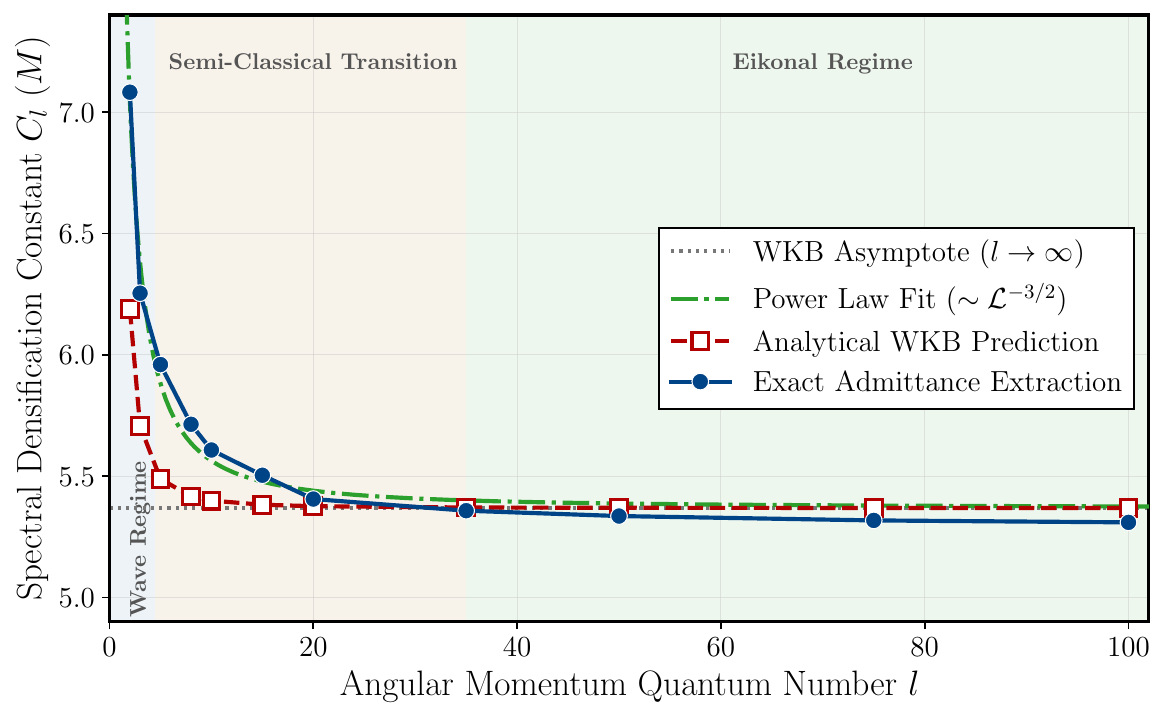}
	\caption{Macroscopic scaling of the gravitational Wigner delay across angular momentum regimes. The densification ranges from a maximal wave-regime delay ($l=2$) down to the effective eikonal asymptote ($l \gg 1$), structurally deviating from the geometric empty-box assumption ($C_l = 0$). The dotted line marks the analytical WKB asymptote, while the dash-dot curve represents the phenomenological inverse power law fit ($\sim \mathcal{L}^{-3/2}$) capturing the macroscopic wave-tunneling deviation.}
	\label{fig:universal_scaling}
\end{figure*}

We evaluate the geometric dependence of the Wigner delay across a range of angular momenta, from the quadrupolar mode ($\ell=2$) up to the deep eikonal regime ($\ell=100$). As demonstrated in Figure~\ref{fig:universal_scaling}, the densification exhibits a distinct structural transition, challenging the empty-box approximation ($\Delta\omega = \pi/L$, corresponding to $C_\ell = 0$) across all observational regimes.

In the eikonal limit ($\ell \gg 1$), the centrifugal barrier dominates the effective potential. The wavefunction becomes highly localized, behaving as a classical light ray grazing the photon sphere, and wave tunneling is suppressed. Consequently, the extracted densification constant approaches the semi-classical WKB asymptote ($C_{\infty} \approx 5.369M$). For $\ell=100$, the numerical evaluation yields $C_{100} = 5.310M$. As appropriately constrained by analytical boundaries, this value falls slightly below the positive-definite WKB limit. In this deep eikonal regime, the physical dispersive delay becomes smaller than the systematic numerical error floor induced by the finite spatial truncation of the solver ($r_{*,\infty} = 200M$). This truncation introduces a residual negative phase shift that mathematically dominates the microscopic physical signal at high multipoles.

In the wave-dominated regime ($\ell \in [2, 10]$), which constitutes the primary emission channel for binary black hole ringdowns, the physics departs from the geometric optics limit. Wave tunneling through the relatively low centrifugal barrier broadens the resonance, causing the wavefunction to flare. This increases the effective interaction time with the spacetime curvature, pushing the densification constant up to $C_2 = 7.082M$. In this domain, the macroscopic phase delay entirely overshadows the numerical truncation error. The deviation from the geometric asymptote is accurately captured by a phenomenological inverse power law governed by the effective angular momentum $\mathcal{L} = \ell + 1/2$:
\begin{equation}
	C_\ell \approx C_{\infty} + \frac{A}{(\ell + 1/2)^{3/2}},
\end{equation}
where $C_{\infty} = 5.369M$ and the wave-coupling constant is fitted to $A \approx 6.625M$. This empirical fit demonstrates excellent agreement ($R^2 > 0.99$) across the wave-tunneling domain. This structural deviation from the semi-classical theory at low angular momenta highlights the necessity of precise phase tracking to systematically capture the full dispersive effects for the most luminous gravitational-wave harmonics.

\subsection{Structural Robustness: Isospectrality and Boundary Phase Invariance}

We ensure the macroscopic densification is not an artifact of idealized assumptions by testing the framework against structural perturbations, a necessary verification given the known pseudospectral instabilities of black hole quasinormal modes \cite{Jaramillo2021}.

\begin{figure}[htbp]
	\centering
	\includegraphics[width=0.85\linewidth]{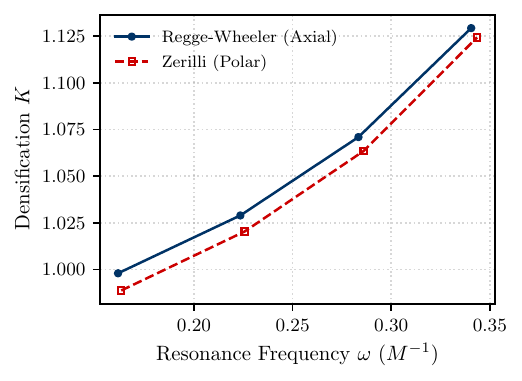} \\
	\vspace{0.6cm}
	\includegraphics[width=0.85\linewidth]{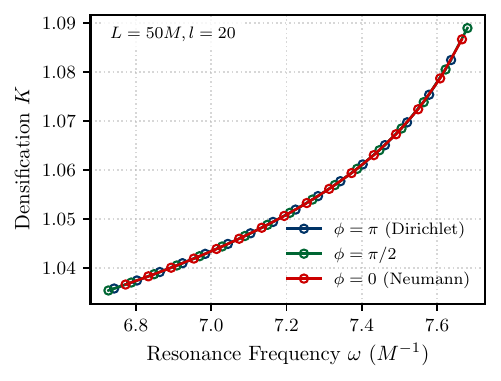}
	\caption{Validation of the structural robustness of the macroscopic phase dispersion. \textbf{Top:} Comparison of the spectral densification $K(\omega)$ extracted from the axial (Regge-Wheeler) and polar (Zerilli) potentials for the quadrupolar mode ($l=2$) at $L=50M$. The relative error remains bounded below $1\%$, confirming the effective isospectrality of the observable despite the finite-boundary symmetry breaking. \textbf{Bottom:} Influence of a static membrane phase shift $\phi$ ($L=50M, l=20$). The exact superposition of the Dirichlet ($\phi=\pi$), Neumann ($\phi=0$), and intermediate ($\phi=\pi/2$) conditions demonstrates invariance against static boundary choices. A genuinely dispersive boundary ($\dd\phi/\dd\omega \neq 0$) would actively modulate this local spacing.}
	\label{fig:robustness}
\end{figure}

First, classical General Relativity dictates that the axial (Regge-Wheeler) and polar (Zerilli) sectors are isospectral in the infinite domain. Introducing a physical membrane at a finite radius formally breaks this symmetry: the underlying Darboux-Chandrasekhar transformation mixes the wavefunction with its spatial derivative, meaning a strict Dirichlet condition ($\Psi = 0$) on the axial sector does not map to a vanishing polar boundary condition. However, our numerical spectral analysis confirms that the axial and polar densification factors $K(\omega)$ remain superimposed. Even in the worst-case wave-tunneling domain ($\ell=2$), the maximal relative error between the Regge-Wheeler and Zerilli macroscopic extractions is bounded by $\mathcal{O}(10^{-2})$ ($< 1\%$), as visually confirmed in the top panel of Figure~\ref{fig:robustness}. The structural difference between the potentials is therefore negligible compared to the dominant macroscopic dispersion, preserving effective isospectrality for the macroscopic WGM comb.

Second, we evaluated the model against complex, static membrane reflectivities $\mathcal{R} = |\mathcal{R}| e^{\ii\phi}$. By adjusting the initial impedance mapping to $Z_{\text{wall}} = k_{\text{wall}} [1 - \mathcal{R}]/[1 + \mathcal{R}]$, we tested three distinct constant phase shift paradigms ($\phi = \pi, \pi/2, 0$). While the absolute resonance frequencies are shifted globally, the densification factor curves overlap identically, as demonstrated in the bottom panel of Figure~\ref{fig:robustness}. The extracted FSR is inversely proportional to the frequency derivative of the total phase; for any static boundary phase shift ($\dd\phi/\dd\omega = 0$), the boundary microphysics vanishes from the derivative. Therefore, the macroscopic densification $C_l$ effectively isolates the structural dispersion of the external spacetime, independent of the ECO surface state, provided the boundary phase remains frequency-independent. As a physical corollary, for genuinely dispersive membranes ($\dd\phi/\dd\omega \neq 0$), this intrinsic boundary delay would couple directly to the local spectral spacing.

	\subsection{Validity of Real-Axis Spectral Extraction}
	
	Whispering Gallery Modes in open black hole potentials are formally leaky resonances defined by complex frequencies $\omega = \omega_R + \ii\omega_I$, where the imaginary part ($\omega_I < 0$) quantifies the decay rate via wave tunneling. We isolate the WGM comb by minimizing the impedance matching residual strictly along the real axis. Analytically, a Taylor expansion of the residual around the complex pole $\omega_c$ demonstrates that the real-axis minimum $\omega_0$ deviates from the true real frequency $\omega_R$ strictly at the second order of the decay rate, yielding a frequency shift $\Delta\omega \propto \mathcal{O}(\omega_I^2)$. 
	
	We confirm this analytical bound by extending the impedance framework to the full complex plane. In the eikonal and semi-classical regimes ($l \ge 10$), the centrifugal barrier effectively suppresses tunneling ($|\omega_I| < 10^{-4}$), rendering the real-axis deviation formally negligible ($\Delta\omega \sim 10^{-6}$). Even in the most extreme wave-tunneling regime ($l=2$) near the barrier peak, where the leakage is maximal ($|\omega_I| \approx 2.9 \times 10^{-3}$), the induced shift on the real frequency remains bounded at $\Delta\omega \approx 3.7 \times 10^{-4}$. This limits the maximum systematic error on the spectral densification factor $K$ to roughly 0.1\%. The real-axis projection is therefore strictly validated, and the extraction error is negligible compared to the macroscopic dispersion governing the hyperbolic scaling laws.
	
	\section{Discussion and Limitations}
	
	While the macroscopic impedance framework formally isolates the dispersive scaling law of the ECO cavity across all angular momentum regimes, its validity is strictly bounded by the following assumptions, precluding direct application to raw interferometric data.
	
	\paragraph{Symmetry Breaking and the Kerr Metric.}
	This study is confined to the static Schwarzschild geometry ($a=0$). Rotation explicitly breaks spherical symmetry. While perturbations in the Kerr metric are separable via the Teukolsky equation, the resulting angular eigenvalues become inherently frequency-dependent ($a\omega$). Mapping this to a real, short-ranged barrier requires the Sasaki-Nakamura transformation, which drastically complicates the definition of a local wavevector. Extending the macroscopic impedance mapping to accommodate this $\omega$-dependent angular mixing remains a formal challenge; thus, the structural invariance of the scaling constant $C_l$ remains to be formally proven for highly spinning ECOs.
	
	\paragraph{The Hollow Cavity Assumption.}
	Our derivation assumes the ECO interior is a vacuum governed by the unperturbed Schwarzschild metric. Physically motivated models (e.g., fluid gravastars) feature interiors filled with exotic matter. Such states modify the effective gravitational refractive index, altering the phase velocity of the wave and breaking the rigid geometric relation $\tau_{\text{geom}} \approx 2L$.
	
	\paragraph{Restriction to Stationary Scattering.}
	The Regge-Wheeler formalism deployed here is a stationary linear perturbation theory. While post-merger gravitational-wave echoes propagate in a linear spacetime regime, the initial wavepacket injecting energy into the cavity is generated by the non-linear binary coalescence. Because this frequency-domain extraction isolates the stationary poles of the $S$-matrix, it identifies the spectral frequencies (the FSR comb) but bypasses the initial transient excitation. The present formalism provides precision templates for the echo phase evolution, but cannot predict their absolute initial amplitudes.

	\section{Conclusion}
	
	This work quantifies the macroscopic phase dispersion governing the spectroscopy of exotic compact objects. By implementing a spectral averaging protocol over the leaky spectral window, we extract the frequency-dependent delays that neutralize the logarithmic singularities of semi-classical theories. This spectral analysis systematically challenges the geometric rigid-box approximation ($\Delta\omega \approx \pi/L$) widely used in current echo template generation. We demonstrate that the resulting hyperbolic densification ($K \sim 1 + C_\ell/L$) accumulates macroscopic phase shifts, and we characterize the structural transition between the wave and eikonal regimes: the deviation from the semi-classical asymptote is dictated by an $\mathcal{L}^{-3/2}$ inverse power law, which captures the signature of macroscopic wave tunneling at low multipoles. Furthermore, this analysis isolates a robust physical observable: the dispersive curvature of the Free Spectral Range is a signature of the external macroscopic spacetime geometry, independent of the reflecting surface state, provided the boundary phase is non-dispersive. By bypassing the limitations of first-order approximations in the deep tunneling regime, this phenomenological framework provides the robust theoretical constraints required to generate high-fidelity echo templates for the next generation of gravitational-wave observatories.

	\section*{Data availability statement}
	The data that support the findings of this study are available upon reasonable request from the authors.
	
	\section*{Conflict of interest}
	The author declares no conflicts of interest.
	
	\section*{Acknowledgments}
	The author acknowledges the use of a Large Language Model (Gemini) for linguistic refinement, proofreading assistance, and LaTeX formatting during the preparation of this manuscript.
	
	\bibliography{references}
	
\end{document}